\documentclass[12 pt]{article}
\usepackage{geometry}                
\usepackage{graphicx}
\usepackage{amssymb}
\usepackage{epstopdf}
\usepackage{url}

\begin{document}
\thispagestyle{empty}

\begin{center}
	\vskip 3cm {\LARGE {\bf Gravity as an Entropic Phenomenon}}\\
	\vskip 1.25 cm  { Abhiram Chivukula } \\
	\emph{Mentor: John H. Schwarz} \\

	\vskip 0.5 cm California Institute of Technology

\vspace{2cm}

\begin{abstract}
\baselineskip=16pt
The unification of gravity with the three other forces has been an important goal of physics for some time now, because a quantum theory of gravity is necessary to explain the universe at its earliest moments. Its pursuit has largely assumed gravity's independent existence, but E. Verlinde proposed that gravity is not a fundamental force but a macroscopic phenomenon that emerges as a result of thermodynamic principles applied to the ÒinformationÓ of mass distributions. Under this framework we consider the roles played by quantum microstates, entanglement, information theory, the AdS/CFT Correspondence, and String Theory in general. We also ask whether Verlinde's proposal suggests that action principles should be thermodynamic in nature.
\end{abstract}
\end{center}

\newpage

\section{Introduction}

	\subsection{Background}
The search for a quantum theory of gravity has been one of the most fundamental problems in physics for the past fifty years because such a theory is necessary to understand the Universe at its earliest moments. Relativistic Field Quantization, having been successfully applied to the Electromagnetic force, seemed to be a natural candidate to quantize gravity, as its fundamental framework had delivered successful results to spectacular accuracies. \\
\indent Subsequently, the GWS model had unified the Electromagnetic and Weak force into a single force with a twofold manifestation due to symmetry-breaking. Attempts to quantize the Strong force initially included String Theory, but Gell-Mann and Fritzsch's Quantum Chromodynamics met more initial success. Several models for Grand Unified Theories then arose, which attempted to combine the Strong and Electroweak forces, though none are yet universally accepted. \\
\indent However, it became evident that the standard relativistic quantization paradigm led to intractable difficulties when applied to Gravity; the perturbative expansions which worked so well when applied to the other fundamental forces instead diverged to give nonsensically infinite answers. The infinite difficulties that beset the then-standard approaches to quantum theories of gravity (let alone a unified theory of gravity) led to String Theory's re-examination. Though it  had been pitched originally as a theory describing the Strong interaction of hadrons, Schwarz and Scherk proposed that it was instead a theory of Gravity, where particles were different oscillatory modes of 1-dimensional ``strings''. Subsequent developments in the 1980s and 1990s positioned String Theory as the most promising candidate for a ``Theory of Everything'', and Maldacena's discovery of the Ads/CFT Duality provided a concrete example of the holographic principle, where a  lower-dimensional theory encodes higher-dimensional physics \cite{Wikipedia}.
  
	\subsection[Motivation] {Motivation \footnote{Paraphrased from Verlinde's ``On the Origin of Gravity and the Laws of Newton''}}
	The copious difficulties with the attempted unification of gravity with quantum mechanics at the Planck scale (that are not all yet understood in a String Theoretic context) led E. Verlinde to propose that perhaps such attempts are misguided. He argues that gravity is not fundamental, but a macroscopic phenomenon that emerges from the thermodynamics of ``information'' in a theory without gravity. Consequently, gravity becomes an entropic force caused by gradients of ``information''.\\
	\indent To arrive at this conclusion Verlinde invokes the applicability of the Holographic principle, and his initial assumptions are that there is some well-defined notion of time; that there are space-independent quantities called Energy, Entropy, and Temperature (which are related by standard thermodynamic relationships); that the number of degrees of freedom associated with a portion of space are finite (required by the holographic principle); that there is an equivalence between energy and matter; and that the energy is distributed evenly over the degrees of freedom in the volume. These assumptions directly lead to an associated temperature for the specified volume, and Verlinde asserts that the product of this Temperature with a differential entropy change (resulting from mass displacement) manifests itself as the gravitational force.\\
	\indent Verlinde elucidates the mechanics of entropic forces in a more familiar setting: the thermodynamics of a polymer immersed in a heat bath. The random collisions from the interaction of the heat bath and the polymer will cause the polymer to probabilistically favor ``curled'' states because there are many more possible curled states than there are ``straight'' states and all such states are equi-energetic. Since the endpoints of the polymer for the curled states must be closer than for straight states, the heat bath immersion tends to create an effective ``restoring force'' that acts to decrease the distance between endpoints. However, this restoring force is a macroscopic phenomenon and is independent of the ``fundamental'' forces governing the actual collisions between the heat bath and the polymer; the restoring force only depends on the statistical mechanics of the system's arrangement. If Gravity is indeed an entropic force then it is in a similar sense not ``fundamental''.\\
	\indent Positing that ``space'' is literally a storage device for mass information, Verlinde suggests that ``screens'' store information about particles that have crossed them, where ``screens'' are lower dimensional manifolds on which some microscopic dynamics are encoded (perhaps a Quantum Field Theory). He quantifies the maximum possible information stored on a screen to be proportional to its area and states that this information is ``processed'' by ``microscopic'' dynamics  that live \emph{only} on the screen. This microscopic theory would be analogous to the forces governing collisions in the polymer example, and its minutiae aren't important to Verlinde's theory; only the overall statistical behavior of it is. The only requirement imposed on the dynamics of the microscopic theory is that it must have a well-defined Energy.\\
	\indent Inspired by Bekenstein's original argument regarding black-hole thermodynamics, Verlinde proposes that the differential change in entropy as a mass crosses a screen is proportional to the mass and the differential displacement: $\Delta S \propto m\, \Delta x$, and invoking the standard thermodynamic relation $F \Delta x = T \Delta S$ we find that $F \propto T$, where $T$ is the temperature of the ``information'' of the mass distribution and $F$ is the force experienced by the mass.\\
	\indent Since the maximum amount of information that can be stored on a screen is proportional to the area, $A$, of the screen (as per the initial quantification), we have $N \propto A$ where $N$ is the number of ``Bits'' of information. Since the system's energy is divided over the $N$ bits, the relation between energy and temperature is $E=\frac{1}{2} N k_B T$ (as per standard thermodynamic relations). We combine this with our initial assumption of the validity of the energy-mass equivalence $E=Mc^2$ where $M$ is the effective mass associated with all the particles that have crossed the screen. Motivated by symmetry, we assume a spherical screen of Radius $R$. Then $A=4 \pi R^2$, and appropriately combining these equations with the $F \propto T$ relation gives $F \propto \frac{Mm}{R^2}$, which is Newton's law of gravitation for appropriate choices of proportionality constants. Thus gravity in this picture is a force that emerges due to the thermodynamics of the information on holographic surfaces.\cite{Verlinde} \\
\indent Verlinde generalizes this example to work with arbitrary matter arrangements (letting holographic screens correspond to equipotential surfaces) and provides a similar argument to heuristically derive the Einstein Field Equations of General Relativity. It is not clear whether Entropic Gravity alleviates prior difficulties with understanding the Universe at its earliest moments.

	\subsection{Results}
	Verlinde's idea recasts gravity as an effective emergent force that only has a meaningful identity at the macroscopic level. This is quite characteristically different from the orthodox view that gravity is one of the fundamental forces of the universe, and is transmitted by a graviton. If Verlinde is correct, then the ``graviton'' is analogous to a phonon in that it is a quantized macroscopic excitation that is non-fundamental.  This perspective shift must lead to new insights and to materially different predictions if it is to be a meaningful theory. By adopting Verlinde's perspective on the nature of gravity we arrive at several questions that beg for further theoretical exploration:\\
	\begin{enumerate}
		\item Should thermodynamics be built into the underpinnings of a theory, or should it be used solely as a analytic tool?
		\item What exactly is the ``information'' to which Verlinde refers? Is its associated entropy intensive or extensive? 
		\item Does the Entropic view of Gravity compel characteristic insights or experimental predictions? 
		\item Verlinde uses ``information'' to make Gravity an emergent phenomenon. Can Quantum Mechanics similarly emerge? 
		\item Can String Theory be reconciled with Entropic Gravity? 
	\end{enumerate}

\subsubsection{Thermodynamic Action}	
\indent For any system that contains a large number of objects we can explore its macroscopic properties by application of statistical mechanics and thermodynamics. This procedure involves identifying a collection of microscopic states with one macroscopic state specifying the system's Energy, Entropy, Temperature, etc. Based on this identification, we can predict how some quantities will change when we vary others under certain constraints. Thermodynamics is traditionally applied to systems in this manner, but if we want a theory of gravity that is thermodynamic in origin and rigorous formulated predictive power, we propose that perhaps that the theory should be written in terms of an action principle that itself contains the thermodynamics. Action principles are routinely used to formulate theories due to the ease with which they allow us to implement constraints and symmetries, and perhaps such an approach would allow the use of standard methods of physical analysis and mathematical techniques to tease out the idea's implications. \\

\subsubsection{Properties of Information}
\indent Though Verlinde quantifies how much ``information'' a particular holographic screen can contain, he doesn't unambiguously define what it actually is. We found that the ``information'' to which Verlinde refers is \emph{not} characterized by the degeneracy of angular momentum quantum states on the holographic surface. Such a characterization would have been intensive, since it wouldn't depend upon the quantity of matter in the system, so we propose that the entropy to which Verlinde refers is extensive with respect to the higher-dimensional space, but intensive with respect to the holographic surface.\\

\subsubsection{Characteristic Insights and Predictions}
\indent Verlinde uses the standard thermodynamic formulation without much precision, but this could pose a problem. The scenarios that would be encompassed by the use of this new approach to gravity include relativistic scenarios. We propose a relativistic reformulation of thermodynamics that would allow rigorously justifiable application to Entropic Gravity.	\\
\indent The entropic formulation of gravity relegates the graviton to being an ``effective'' macroscopic quantum, analogous to a phonon \cite{Verlindeblog}. Extending this analogue, we predict short range properties of the phonon that possibly extend the use of quantum field theoretic computational methods to gravity at the Planck scale by assigning it an effective mass.\\

\subsubsection{String Theory and Emergent Quantum Mechanics}
\indent By starting from first principles, information, and thermodynamics Verlinde derives Newtonian and Einsteinian gravity. We propose that Quantum Mechanics similarly emerges from information and justify this proposition with an information-based heuristic derivation of Heisenberg's Uncertainty Principle. An emergent Quantum Mechanics is further justified by the dualities of string theory, which inextricably intertwine large scale and small scale effects. We quantify the heuristic properties that such an emergent quantum mechanics must satisfy if it is necessitated by these dualities, and we propose a possible framework outlining the emergence of Quantum Mechanics through ``Markov-like'' chains and modifications of the fundamental principles of statistical mechanics.\\


\section{Methodology}
\indent We present the important points of interest that influenced our thought progression to arrive at our conclusions. We first present influential details of Verlinde's approach to entropic gravity, and we justify the resultant proposals made above. 
	\subsection{Influential Points and Discussion}
		\subsubsection{Assumptions and Framework}
		\indent Verlinde points out the similarity between gravitational, thermodynamical, and hydrodynamical laws, but his paper addresses only the similarity between gravity and thermodynamics. Son and Starinets \cite{Sonstar} discuss the link between gravity and hydrodynamics, but Verlinde's discussion doesn't elevate the role of hydrodynamics to that of thermodynamics (in its relation to gravity), since he argues that gravity follows from thermodynamic principles but makes no comment regarding the importance of the relationship between hydrodynamics and gravity (other than saying that it implies a certain universality).\\
\indent Gravity and space-time geometry are ``emergent phenomena'' in Verlinde's framework, implying that they have no fundamental microscopic definition, but arise as macroscopic behavior. He cites the AdS/CFT correspondence as an example of how a non-gravitational theory can give rise to a theory of gravity to justify his argument for gravity's emergence.\\
\indent To further justify gravity's non-fundamentality, Verlinde argues that gravity's interaction with all fields (as prescribed by General Relativity's requirement that anything with energy influences and responds to gravity) suggests that its mechanism is independent of any particular field theory's specifics. Verlinde defines ``information'' to depend on the amount and distribution of matter (measured in terms of the entropy of the ``microscopic theory'') and asserts its thermodynamics manifests itself as gravity.\\
\indent That information associated with space obeys the holographic principle (justified by black hole physics and the AdS/CFT Correspondence) is a fundamental assumption that Verlinde makes. He discusses emergence of space and gravity through the holographic principle, along with a coarse graining procedure, arguing that if space itself is emergent, then Newton's laws have to be derived because concepts such as position, velocity, and acceleration,  are no longer intuitive.\\
\indent From space-independent concepts such as energy, entropy, and temperature Verlinde derives Newton's laws, and Gravity is explained as an entropic force caused by a change in the amount of information associated with the positions of bodies of matter. Verlinde's assumptions here are somewhat worrisome because though energy, entropy, and temperature are space-independent, they are not time-independent; in fact, having an energy requires a well-defined notion of time. This is problematic because Relativity treats space and time on equal footings and inextricably intertwines their behavior. Relativity's countless experimental confirmations would require of Verlinde's theory a preponderance of predictive utility to once again separate time from space. Furthermore, this requires a reconciliation between the grounding fundamentals of Verlinde's theory and Relativity because in general (relativity) we do not  have time-like Killing vectors in arbitrary spacetimes, but Verlinde \emph{requires} a time-like Killing vector in his relativistic derivation. It may be difficult to define dynamics without giving time a special footing, but human conceptual difficulties with physics are not enough justification for altering fundamental assumptions. \\
\indent The validity of the Holographic Principle requires that a finite number of degrees of freedom are associated with a given spatial volume. This is exemplified by the ``discrete-chunk'' nature of the polymer chain, so the mechanical picture provided to elucidate the character of entropic forces is a fair comparison. The distribution of mass-energy over all the degrees of freedom (justified by equipartition of energy) is another assumption that Verlinde makes, but (as he himself notes) it is not crucial that it is strictly obeyed as long as the number of bits of information is large enough so that the central limit theorem is applicable. This distribution gives a characteristic temperature, and with this temperature he equates the product of a temperature and the entropy change with that of the net force and displacement to relate force to temperature. Those steps only require the validity of simple statistical mechanics and thermodynamical arguments, which in the non-relativistic case is fine, but we contend that the validity of such arguments is less clear when dealing with the relativistic case due to the less well defined nature of energy of macroscopic systems in General Relativity. \\

\subsubsection{Force and Inertia}

Verlinde replicates Bekenstein's thought experiment, which treats matter at one Compton wavelength from an event horizon to be part of the associated black hole. In Verlinde's case the black hole is replaced with a holographc screen, and at one Compton wavelength mass and horizon area increases by one ``bit's'' worth. Verlinde reproduces this argument in flat space next to a holographic screen, taking the change in entropy to be $\Delta S=2 \pi k_B$ for $\Delta x =\hbar/mc$, which assumes that mass and entropy are additive, and to get an associated force he assumes  $F \Delta x=T \Delta S$. We note that this implies a Galilean force transformation law, since there is no relativistic velocity correction term in $F \Delta x=T \Delta S$. As this derivation assumes non-relativistic velocities, this aside has no ad-hoc impact on its validity, but if one wished to provide a better relativistic derivation than Verlinde's, modifying the additivity of mass, and the entropic force experienced at a certain temperature would be one approach to doing it. The previous relations are combined to give: 

\[\frac{\Delta S}{ \Delta x}=2\pi k_B \frac{mc}{\hbar}\]
\[F \Delta x=T \Delta S\]
\[F \Delta x=2\pi T  \frac{mc}{\hbar} k_B \Delta x\]
\[F =2\pi  \frac{mc}{\hbar} k_B T\]

\indent Verlinde arrives at the conclusion that at an ``information'' temperature $T$, a mass $m$ near a the holographic surface experiences a force $F=2\pi  \frac{mc}{\hbar}  k_B T$. He subsequently appeals to the Unruh Effect, given by, $k_BT=\frac{\hbar a}{2 \pi c}$ (where $a$ denotes acceleration) to ``derive'' $F=ma$. 

\indent The Unruh Effect is a phenomenon that causes accelerated observers to perceive quantum fields to be in excited finite temperature states, whereas uniformly moving observers perceive the same fields to be in vacuum states. Therefore, each observer measures a different particle number, and the accelerated observers experience an associated temperature of space. However, Verlinde interprets the Unruh Effect equation differently, suggesting that putting the system at temperature $T$ produces an acceleration $a$. This interpretation is not simply a demonstration of time reversal symmetry, as the time reversed Unruh effect is \emph{still} the Unruh effect. We contend that it is actually a thermodynamic reversal, much like the relationship between a heat pump and an engine. Importantly, we note that the radiation is \emph{not} isotropic, as it plays a role in illustrating the mechanism by which this ``reversal'' occurs.\\ 
\indent To explain this mechanism we look at an analogue: A blackbody at an (ordinary) temperature $T$ emits photons according to the blackbody spectrum, whereas an object at an information-temperature $T$ emits graviton quasi-particles according to a blackbody spectrum. These gravitons are not fundamental particles; rather, they are analogous to phonons, and represent propagating quantized rarefactions of information-entropy gradients.  Instead of acceleration producing particles, as is the case with the Unruh Effect, these particles produce the acceleration experienced. This acceleration causes particles with mass to cross the holographic screens, which causes the information temperature on those screens to increase. This increased information temperature causes an increased acceleration for particles outside the screen. This picture is plausible because the graviton is also massless and its non-relativistic similarity in form to the Coulombic force law. Therefore the idea that an object with an information temperature similarly information-radiates a blackbody spectrum in the form of phonon-like gravitons is consistent with the entropic gravity framework. \\
\indent Temperature is a measure of internal energy, and internal energy that couples to the electromagnetic field is radiated away in the form of light. Similarly, the internal energy that couples to ``information'' (or information temperature), manifests as entropic gravity, with a phonon-like quasi-particulate graviton. We commonly call this internal energy that couples to ``information'' by the name ``Mass'', to which the thermal temperature clearly contributes.  To start off his derivation Verlinde assumes that the maximum possible information, $N$, contained on a screen is proportional to the area, $A$, of the screen. He then chooses fundamental constants to nondimensionalize appropriately: $N = \frac{Ac^3}{G \hbar}$.The rest of Verlinde's derivation for the Newtonian gravitational force law is as follows:

\[N=\frac{Ac^3}{G \hbar}\]
\[E=\frac{1}{2} N k_b T\]
\[E=Mc^2\]
\[Mc^2=\frac{1}{2} N k_B T\]
\[F =2\pi  \frac{mc}{\hbar} k_B T \rightarrow \frac{F \hbar}{2 \pi mc}=k_B T  \]
\[Mc^2=\frac{1}{2}N  \frac{F \hbar}{2 \pi mc}\]
\[Mc^2=\frac{1}{2} \left( \frac{Ac^3}{G \hbar} \right)  \frac{F \hbar}{2 \pi mc}\]
\[M=\left( \frac{A}{G } \right)  \frac{F }{4 \pi m}\]
\[M=\left( \frac{4 \pi r^2}{G } \right)  \frac{F }{4 \pi m}\]
\[F=G\frac{Mm}{r^2}\]

\subsubsection{Robustness of the Derivation}

Verlinde states that the starting point in his argument was that space has ``one emergent holographic direction'' but his discussion of this idea is unclear. His claim that space doesn't exist on one side of a holographic membrane is somewhat foreign and at odds with the view in General Relativity, where even though a far away observer never sees any object reach the event horizon of a Schwarzschild black hole, there isn't a true singularity there, as one can simply change from Schwarzchild coordinates to Eddington-Finkelstein coordinates to show the illusory nature of the apparent divergence. Perhaps Verlinde means that there is some evolving coordinate system in which the coordinates are unbounded, but the metric is such that integrating over its volume element gives a finite volume, giving an effective asymptotic surface. Supposing this point were clarified and formalized, we could better illustrate the foliation and coarse graining of holographic screens as space itself emerges. \\
\indent Verlinde states that the defining assumptions are the entropic behavior, the degrees of freedom being proportional to the screen area, and the equipartition of energy over these degrees. He also discusses the necessity to include the mass-energy equivalence and says one cannot neglect Relativity, but this is really not strictly a Relativistic contribution (if we define a ``Relativistic contribution'' as a term that becomes important at large velocities), as there is really no issue with adding an arbitrary constant to the hamiltonian under Newtonian Mechanics  (the $mc^2$ mass energy term). Therefore relativity is really not necessary in the Newtonian discussion, and we just treat $c$ as a conversion factor for units. We clarify this point; the $mc^2$ term in the relativistic energy expansion is zeroth order in $v$, so it is clearly a valid rest energy for non-relativistic scenarios. An example of where this might occur is if mass were converted to energy, and an aggregate mass recoiled via the Mossbauer Effect. The recoil velocity would be non-relativistic for such a large mass. The idea that mass and energy can be interchanged was predicted by Relativity, but it did not have to be unique to relativity. \\
\indent Verlinde then includes a discussion on the necessity of $\hbar$ in the derivation, and he concludes that $\hbar$ is really just an auxiliary variable. The relegation of $\hbar$ to a passive participant is somewhat unfair because these derivations could just be the zeroth order effects, and there could be higher order quantum corrections. If we accept Verlinde's argument that ``the central notion needed to derive gravity is information'," then if that information (and information theory) is intimately connected with quantum effects it may not be so easy to separate an emergent Quantum Mechanics from Statistical Mechanics/Thermodynamics and ``information''. We contend that there is perhaps a deeper connection between quantum mechanics and information, and that the ``auxiliary'' role of $\hbar$ is due possibly to a vanishing of the first order effect.\\

	\subsection{Thermodynamic Action }
\indent Verlinde's idea in some sense is reminiscent of the Bohr Model in that it aggregates a disparate collection of ideas (Holography, Entropy, Energy-Mass Equivalence, and Equipartition), while seemingly striking a grain of truth. Van Raamsdonk's \cite{Raamsdonk}  ideas on entanglement-entropy and mutual information are alluring due to the physical picture of connecting space they provide, but retroactively adding it into Verlinde's framework would make it seem increasingly haphazard. We propose that these ideas be coherently merged together via an action principle.\\
\indent Since Verlinde's ideas are so thermodynamic in nature, this action principle itself should contain the thermodynamics. It seems like Verlinde himself hints at this when he restates the thermodynamical equations in terms of an extremal principle. Furthermore, when he concludes with this principle in (5.31), he states in the paragraph after that he should have really gone backwards. \cite{Verlinde} \\
\indent To state an action principle that incorporates both quantum entanglement, thermodynamics, and the AdS/CFT Correspondence, the Von Neumann Entropy is a natural concept with which to proceed. In the spirit of Verlinde's approach in equation (5.31) the action principle should be similar to ``The Von Neumann Entropy should be extremal with respect to Density Matrix of the Quantum Field (where space has emerged) for a fixed energy'': 

\[ \rho_A =Tr_{\bar{A}} \left( \vert \Psi \rangle  \langle \Psi \vert \right)\] 
\[S =- Tr \left( \rho_A \log \rho_A \right) \]
\[ \delta S \left(\rho_A, E  \right)=0 \]

With the appropriate restrictions on the Trace of the Density Matrix (using Lagrange Multipliers), we get the following:

\[ S=  Tr \left( \rho_A \log \rho_A \right) + \alpha \left( Tr \left( \rho_A \right) -1 \right) +\beta \left( Tr \left( \rho_A H \right) -E\right) \]

\indent This gives the action (under variation of the quantum field's density matrix) to find the Thermal Density Matrix, as Van Raamsdonk \cite{Raamsdonk} references on page 20 and in footnote 19, which means that the perturbing force term that pulls the system out of equilibrium must be added. The ``External Force Term'' in the polymer example was $Fx$ and in the relativistic case it was $ e^{\phi (x)} m $.  It seemed like since Verlinde was  using the ``polymer model'', this gave ``Hooke's Law'' and so the $Fx$ term is supposed to represent the energy put into the system by pulling the polymer out of equilibrium. It is clear that some modification needs to be made to the action to account for this external force, and based on the AdS/CFT correspondence it should be an invariant term that characterizes gravity in String or String Field Theory. We want the theory to be Lorentz-invariant, so instead of an ``Energy'' term, we should add some invariant; we propose that this invariant be the Nambu-Goto action for a closed String.\\

	\subsection{Properties of Information}
	Verlinde posits that space is ``literally just a storage space for information'' and that the maximal allowed information is finite for each part of space, but it is not clear exactly what Verlinde means by these ``bits''. We therefore explore quantum mechanics on the holographic surfaces to better understand what ``bits'' he is talking about.\\
		\subsubsection{Bit Quantization}
\indent  In section 3.2 Verlinde makes the assumption that the number of bits stored on a holographic surface is proportional to the surface's area, $\frac{Ac^3}{G\hbar}$. It would seem to be a reasonable assumption, as it assumes some spatial symmetry of information under the idea of holography, but in what manner would these bits be stored? Our previous discussions on the seemingly fundamental connection between quantum mechanics and information motivates us to search the quantum realm for clues as to the nature of the assumption that $N=\frac{Ac^3}{G \hbar}$. For perspective: under this assumption, 1 bit of information is packed so densely that it would be stored on a sphere with a radius equal $10^{-35} \rm{m}$.\\
\indent We first consider the quantum system of 1 particle trapped in a spherical shell. We set up Schr$\rm{\ddot{o}}$dinger's equation in Spherical Coordinates and simply discard the radial component of the Laplacian:

\[-\frac{\hbar^2}{2m} \nabla^2 \psi = E \psi\]
\[ \nabla^2_{sphere} \psi = -k^2 \psi\]

Using separation of variables we find that our solutions are the spherical harmonics and the complex exponential with typical quantization of angular momentum.

\[\psi =\Phi \left( \phi \right) \Theta \left(\theta \right) \] 
\[\Theta \left( \theta \right)=\pm e^{i m \theta}\]
\[\Phi \left( \phi \right)=P^m_l \left( \cos {\theta} \right) \]

With:
\[A=4 \pi r^2\]
\[ -l \le m \le l \longrightarrow N=2l+1\] 
\[l(l+1)=\frac{2mEr^2}{\hbar^2}\]
\[ l= -\frac{1}{2} \pm \frac{1}{2}\sqrt{1+ \frac{8mEr^2}{\hbar^2} }\]
\[N=2l+1=\sqrt{1+2\frac{mEA}{\pi \hbar^2} }\]
\[N=2l+1=\sqrt{1+k^2\frac{A}{\pi} }\]

So for large E or large A, $N \propto A^{1/2} $, which is not the relation we want.
Noticing the factors of $c^3$ we instead consider the relativistic analogue of this equation in the hopes that it produces the proper relation. We start with the relativistic momentum energy relation:

\[E^2=p^2c^2+m^2c^4\]
\[p^2=\frac{E^2-m^2c^4}{c^2}\]
\[\hat{p}^2 \psi = \frac{E^2-m^2c^4}{c^2} \psi\]
\[-\hbar^2 \nabla^2 \psi = \frac{E^2-m^2c^4}{c^2} \psi\]
\[ \nabla^2 \psi = -\frac{E^2-m^2c^4}{\hbar^2c^2} \psi\]
\[ \nabla^2 \psi = -k^2\psi\]

We similarly remove the radial degree of freedom:
\[ \nabla^2_{sphere} \psi = -k_{rel}^2\psi\]

Since we have the equation in the same form as in the non-relativistic case:

\[N= \sqrt{1+k_{rel}^2 \frac{A}{\pi} }\]
\[N \approx \frac{E}{\hbar c} \left(\frac{A}{\pi}\right)^{1/2} \]

So we still get an incorrect number of states. Based on the presence of $G$,  we consider that we may have to generalize the equation to account for gravity. To do this we first write out the time dependent version of the relativistic generalization, taking $p \rightarrow -i \hbar \nabla$ and $E \rightarrow i \hbar \frac{d}{dt}$ :

\[(i\hbar \frac{d}{dt})^2 \psi= c^2\left(-i \hbar \nabla\right)^2 \psi +m^2c^4 \psi\]
\[-\hbar^2 \frac{d^2}{dt^2} \psi= - \hbar^2 c^2 \nabla^2 \psi +m^2c^4 \psi\]
\[- \frac{d^2}{c^2dt^2} \psi= -  \nabla^2 \psi +\frac{m^2c^4}{c^2\hbar^2} \psi\]
\[- \frac{d^2}{c^2dt^2} \psi + \nabla^2 \psi = \frac{m^2c^4}{c^2\hbar^2} \psi\]
\[ \eta^{\mu \nu} \partial_{\mu}\partial_{\nu} \psi =  \frac{m^2c^4}{c^2\hbar^2} \psi  \]

We convert it to it's general relativistic form by changing the metric and replacing the partials with covariant derivatives. Since $\psi$ is a scalar, the covariant derivatives are just partials:

\[ g^{\mu \nu} \nabla_{\mu} \nabla_{\nu} \psi =  \frac{m^2c^2}{\hbar^2} \psi  \]

Since we need to introduce $G$ into this in some manner, so we consider a spherically symmetric metric to uniformly confine the information to the surface. The Schwarzschild metric comes to mind but it has the obvious coordinate singularity at $r_s=2GM/c^2$ so we change to the Lemaitre Metric, which has the same physics but with different coordinates:

\[ds^2=d\tau^2-\frac{r_s}{r} d\rho^2-	r^2 \left( d\theta^2+\sin^2 {\theta} d \phi^2 \right)\]
Where $r=\left(\frac{3}{2} \left( \rho - \tau \right)\right)^{2/3} r^{1/3}_s $

If we wish to fix the coordinates at the Schwarzschild radius, we need to take $dr=0$ and $\frac{3}{2}\left( \rho -\tau \right)=r_s$, which implies that $d\rho = d\tau$, giving:

\[ds^2=d\tau^2-\frac{r_s}{r_s} d\tau^2-	r_s^2 \left( d\theta^2+\sin^2 {\theta} d \phi^2 \right)\]

\[ds^2=r_s^2 \left( d\theta^2+\sin^2 {\theta} d \phi^2 \right)\]
Which is unfortunately not significantly different than before, so we do not expect significantly different behavior from that which we got from the Klein-Gordon equation.

\indent In all three cases, we see that the angular momentum eigenstates under the Klein-Gordon Equation do not seem to represent the ``information'' to which Verlinde is referring, as the information they would encode would not depend linearly on the area.\\ 
\indent We found that the ``bits'' do not seem to be at all related to the angular momentum eigenstates on a holographic surface. However, the bits are related to how much mass has crossed a particular holographic screen, so when considered in the bulk, the entropy seems to be extensive, as it depends on the quantity of the mass.\\ 
\indent However, the dynamics corresponding to the information change that occurs as a mass is crossing a particular holographic would have been initially encoded on the holographic surface. The bulk picture is simply supplementary. In that sense, when we consider the entropy associated with a particular holographic screen, since all the information is encoded on that screen the entropy is certainly intensive, and it simply evolves with time. That evolution of time does not constitute extensive entropy because that evolution was certainly determined by the information on the screen in the first place.\\

	\subsection{Characteristic Insights and Predictions}
	
		\subsubsection{Relativistic Thermodynamics}
		Verlinde uses the standard formulation of thermodynamics without much discretion. In the Newtonian gravity case this is not so much of an issue, but Verlinde severely restricted his relativistic argument so that standard thermodynamics would apply. His assumption of a time-like Killing vector is a stringent constraint that does not hold for arbitrary spacetime, so a well defined energy does not always exist, thus a well defined notion of temperature does not always exist. We also encountered a relativistic issue when we discussed how Verlinde's Newtonian gravity derivation assumes a Galilean entropic force transformation law with respect to velocity. In hopes of generalizing the applicability of the thermodynamic framework we propose a starting point for formulating relativistically invariant thermodynamics: \\
\indent The approach we take is to first formulate the special relativistic case, to put it into covariant form, and to then use the principle of minimal coupling to obtain the equations for the General Relativistic case. For a given system, we define the Proper Entropy $\sigma$ to be the natural log of the Invariant Degeneracy $\omega$, which is an integral over the system's Lorentz-Invariant Phase Space ($LIPS$):
\[\omega=\int dLIPS\]
\[\sigma=\log{\omega}\]  
\indent In the case of a macroscopic number of particles, we shall replace the finite dimensional $dLIPS$ volume element by a continuous product $dLIPS$ volume element and a phase space density. This can be calculated in the system's center of mass frame to ease computation.  As $\omega$ was chosen to be a Lorentz-Invariant quantity, so is $\sigma$.\\
\indent Ordinarily, we would define a temperature here using $T^{-1}=\frac{\partial \sigma}{\partial E}$, with $E$ the energy, but neither $E$ nor $T$ would then be a relativistically invariant quantity. If we accept that physics is geometry, then the laws of physics should only be defined based on meaningful invariant quantities, we instead define the Invariant Temperature $\mathcal{T}$ as such:
\[\mathcal{T}^{-1}=\frac{\partial \sigma}{\partial m}\] 

\indent Since $m$ is the invariant mass, we truly have an Invariant Temperature, which clearly satisfies the requirement that absolute zero be absolute zero in any inertial reference frame. Is it valid that we use the invariant mass $m$ rather than $E$? We argue that it is  valid, as the invariant mass is a measure of the internal energy of the system and when we ordinarily define $T^{-1}=\frac{\partial S}{\partial E}$, we really consider \emph{only} the change in the internal energy. The way we ordinarly define temperature \emph{rightfully} ignores an energy change due to changing the momentum of the center of mass of the system because changing reference frames doesn't contribute to the system's disorder. We rewrite the invariant temperature in covariant form:

  \[\mathcal{T}^{-1}=\frac{\partial \sigma}{\partial X^{\mu} } \frac{\partial X^{\mu}}{\partial m}=\frac{\partial X^{\mu}}{\partial m} \partial_{\mu} \sigma =\eta_{\mu \nu} \frac{\partial X^{\mu}}{\partial m} \partial^{\nu} \sigma\]

\indent We use the principle of minimal coupling and simply replace the metric. Ordinarily we would have to convert the partial to a covariant derivative, but since $\sigma$ is a scalar the covariant derivative is a partial derivative.

  \[\mathcal{T}^{-1}=g_{\mu \nu}  \frac{\partial X^{\mu}}{\partial m}  \partial^{\nu} \sigma\]
  
\indent However, we must note that the invariant degeneracy $\omega$ must now be computed with respect to a differential General-Covariant Phase Space volume element. We notice that in order for this  quantity to be an invariant in both space and time for the General Relativistc case, we must actually take $\sigma$ to be an entropy density $\rho$, and integrate over the entire General-Covariant Phase Space (with volume element $\ -g \, d^4x \, d^4p$).

  \[\mathcal{T}^{-1}=\int -g \, d^4x d^4p \,\,g_{\mu \nu}  \frac{\partial X^{\mu}}{\partial m}  \partial^{\nu}\rho \]
  
		\subsubsection{Extracting Quantum Gravity Predictions Via Phenomenological Effective Field Theory}
\indent In the ``On the Origin of Gravity and the Laws of Newton'', Verlinde suggests that gravity is not a fundamental force, but a consequence of thermodynamics. Pursuing this idea from a string theoretic perspective, Verlinde argues that closed strings (which ordinarily have vibrational modes corresponding to the graviton) are artifacts that we must integrate out, from the viewpoint of entropic gravity.\\
\indent When asked what role gravitons play in entropic gravity, Verlinde responded on his blog \cite{Verlindeblog} that, ``Gravitons are like phonons...[they] exist as `quasi'-particles. But they do not exist as fundamental particles. ''\\
\indent If we seriously consider the graviton to be a phonon analogue, it has surprising and possibly useful consequences; it leads to a phenomenological characterization of space-time behavior that suggests an approach to constructing an Effective Field Theory valid at and below the Planck Length.\\
\indent Phonons are vibrational quanta on a discrete lattice, and Verlinde's comparison of gravitons to phonons naturally invites the existence of some sort of discrete lattice. One might guess that this lattice corresponds to discretized information/space-time in some manner, but the details of this correspondence aren't important for the phenomenology that this discretization induces. If we assume this discretization extends in at least 3 dimensions, we first claim that phonons can be emitted both longitudinally and transversely. \\
\indent Let us assume the dynamics of longitudinal and transverse phonons corresponds roughly to coupled spring systems. For the longitudinal mode, we can easily see that as the phonon wavelength approaches the ``lattice spacing'', the lattice can act as a low-pass filter and attenuate high frequencies, or short wavelengths. In this ``reactive range'' the dispersion relation takes the form of an exponentially decaying amplitude as a function of distance. \\
\indent In the case of transverse waves, we can treat this with a ``beaded spring" model, in which case we similarly have a low-pass filter with the dispersion relation $\omega^2 = \frac{4T_0}{Ma}\sin^2{\frac{\pi a}{\lambda}}$, which implies no propagation for infinite wavelengths as well as a discrete set of very short wavelengths.\\ 
\indent Taking our lattice spacing to be roughly Planck's Length, we'd find our first non propagating wavelength to be Planck's length. As there is large variation in propagation velocities at very short wavelengths, we'd find initially strong oscillations to decohere and therefore diminish rather quickly. We expect to get a similar type of ``exponential decay of amplitude as a function of distance from the radiating point''. If this holds, then we choose a source and we hit it with an impulse and solve for the Green's function. We Fourier expand the Green's function and discard all terms that correspond to wavelengths larger than the Planck length.  We then inverse Fourier Transform to get the decay dynamics of the sub-Planck length oscillations, which is exactly what we wanted, as we can then see the change in amplitude as a function of distance and time. \\
\indent A quasi-exponential type of amplitude decay would suggest the possibility of approximating sub-Planck length interactions with an effective field theory with a massive (necessarily spin-2) graviton. As massive gauge-bosons have finite lifetimes, and thus demonstrate exponential decay over distances, we could assign an ``effective mass" to the graviton for shorter distance scales. At these incredibly short distance scales, the non-lightspeed propagation of the graviton on these scales would contribute negligible error.\\
\indent We suggest that perhaps the spin-2 field with an ``effective mass'' might allow the use of different perturbative tools to approximate the effects of quantum gravity at sub-Planckian length scales, and might produce non-infinite answers.\\

	\subsection{Emergent Quantum Mechanics and String Theory}
	
	Verlinde's proposal that gravity is an entropic phenomenon may imply that quantum mechanics is also an emergent phenomenon. Given that the character of the entropy to which Verlinde refers is string theoretic, this opens up possible lines of reasoning in this framework. Specifically, the S, T, and U-Dualities of string theory couple strong behavior in one formulation of string theory to weak behavior in another. These dualities suggest that if gravity is emergent, then from the perspective of string theory, Quantum Mechanics should be emergent in its dual theory. We first present a heuristic derivation for justification by example, and then provide a framework for a truly emergent quantum mechanics.
	
		\subsubsection{A Heuristic Information-Based Derivation of Heisenberg's Uncertainty Principle}
		Verlinde posits that space is literally storage for phase-space information such that each part of space has a finite information maximum. This interpretation of the nature of space provides an elegant picture of the Heisenberg Uncertainty Principle. We can picture this ``information space'' as ``memory'' where each ``bit'' of memory has an ``address'' and can store ``data''. In this example let the addresses correspond to coordinates in position space and let the data correspond to coordinates in momentum space.\\	
\indent Information in this framework loosely corresponds to knowing with certainty where in phase-space a particle is \emph{not}. If a particle's position is specified particularly well, then it is associated with very few memory addresses (since the copious position \emph{information} occupies the memory addresses of where the particle is \emph{not} located). Consequently, the few remaining addresses can store very little information about the momentum of the particle in their associated memory.\\
\indent Suppose a particle of mass $m$ has a position to precision $\Delta x$, and assume via the ergodic hypothesis an equiprobable distribution within the region $\Delta x$ spans. To compute the associated information (assuming information roughly obeys the entropy formula $I=-S=-log(\Omega)$) we take the logarithm of the degeneracy, which is $\Delta x$ (before suitable non-dimensionalization of units). We restate Verlinde's ``finite information'' condition as, ``A region's net information has a finite upper bound'' to clarify and interpret information as a relative quantity to allow for negative information. Anticipating our final result, we non-dimensionalize our degeneracy, wisely choosing $I_{max}=0$, $\lambda =\hbar/(\sqrt{2}mc)$, and $\mu=mc/\sqrt{2}$ . If we saturate the information bound with $\Delta x$ and $\Delta p$, then $I_{max}=I_{x}+I_{p}$, but supposing $\Delta x' \ge \Delta x$ and $\Delta p' \ge \Delta p$:

 \[I_{max}\ge I_{x'}+I_{p'}\]  
 \[I_{x'}=-log\left( \lambda^{-1} \Delta x' \right)\] 
 \[I_{p'}=-log\left(\mu^{-1} \Delta p' \right)\] 
\[ \Delta x' \Delta p' \ge \lambda \mu e^{-I_{max}}\]

Our choices of $\lambda, \mu,$ and $I_{max}$ give $ \Delta x' \Delta p' \ge \hbar/2$. Though it is interesting that uncertainty decays exponentially as information increases, this result exemplifies the idea that Quantum Mechanics could also be an Entropic Phenomenon.

		\subsubsection{Formalizing the Emergence of Quantum Mechanics}
		
		 To speculate about what an emergent Quantum Mechanics looks like, we must consider how the assumptions that are put into Verlinde's theory translate into Quantum Mechanical assumptions under the duality transforms in String Theory.

O. Penrose explains the postulates of statistical mechanics \cite{Penrose}:

\begin{itemize}
	\item \textbf{1) Dynamics:} The dynamics of systems are governed by quantum or classical mechanics
	\item \textbf{2) Measurement:} Macroscopic systems have variables, about which we can only measure ``indicators'' which take the discrete values of 0 and 1, and such measurements can only be made in discretely spaced units of time.
	\item \textbf{3) Observer Effect:} The disturbance caused by measurement is negligible
	\item \textbf{4) Markovian Postulate}: The successive states of the system constitute a Markov Chain
	\item \textbf{5) Bose/Fermi Symmetry:} There are no artificial restrictions on the dynamical states apart from the Bose and Fermi Symmetry conditions
\end{itemize}

\indent From these postulates he goes on to derive statistical mechanics and thermodynamics. If we want to make quantum mechanics emergent, then we have to modify these postulates themselves, since they assume quantum mechanics as a prerequisite.\\
\indent It is obvious that postulates 1,2, and 5 are directly quantum mechanical, which immediately draws them out as candidates for ``elimination'' as postulates if Quantum Mechanics is to be emergent. Postulate 3 only describes the ``size" of systems, so it isn't relevant.\\
\indent Postulate 4, however, is interesting because it is neither classical nor quantum in its statement. If we imagine representing a Markov Chain in the form of a graph, with transition probabilities from one node to another, then this suggests a malleable picture for classicization or quantization of a large system. With a classical system, we'd simply view different probabilities for transitions from one state to another, assuming that there are well defined values of variables at each node of the Markov Chain. If we wanted to instead make quantum mechanics emerge from this, we only need to throw in a qualifier to the Markov Chain picture, stating that the the amplitude for a transition between any two nodes $a$ and $b$ is the sum over transition amplitudes between all possible discrete paths across the graphical representation of the Markov Chain. This invokes the Feynman Path Integral formulation of Quantum Mechanics, but in terms of justifying this claim, it only needs a slight extension of the purely ``probablistic'' idea that if something can happen in multiple discrete ways then the probability of it happening is the sum of probabilities for each of these methods, which is not so large a jump conceptually.\\
\indent This generalization of  ``classical probability'' may require invoking ideas from noncommutative probability theory. Some assumptions must be input because even in Verlinde's formulation we didn't get something for nothing; he made assumptions about the way ``information'' behaves.  Specifically, he equi-partitions the energy of the system over the bits which we posit roughly translates to treating each path with an equal amplitude in this ``emergent'' picture. Following this assumption all other "postulates" follow as consequences.\\
\indent What further suggests that this may be on the right track to making quantum mechanics emergent is that the Partition Function itself looks very much like a Feynman Path Integral. Since the similarity is so striking we could possibly absorb the factor of $i$ and actually call it a partition function and phrase everything in terms of thermodynamical statements. 
		
		\subsubsection{Relationship to String Theory}
		Though we have suggested many different approaches for furthering Verlinde's ideas. None of these ideas are the manner in which he is exploring how to further the theory. In his talk at Caltech Verlinde suggested modifying the nature of phase-space itself by varying the width of a supersymmetric harmonic oscillator potential at each point in space. Even though there is no potential gradient along the direction of the increase in width the object still experiences an effective force towards the direction of increasing phase space size. Verlinde, however, takes his entropic force to be adiabatic, and using the Born-Oppenheimer approximation, creates ``slow'' and ''fast'' variables. Verlinde's ``fast'' variables encode symmetries and all interactions of string matrices. By contrast, his slow variables only contain the diagonals of the string position matrix, which for large distances relative to the string-length coincides rather well with the matrix's eigenvalues, and therefore the the notion of ``distance''. When these strings become too close, the coincidence is lost due to magnification of the off diagonal terms and the higher order corrections require the fast variables. Verlinde uses the degeneracy of the string matrix's off diagonal terms to generate the size of the phase at a holographic screen.

\section{Conclusions}
	We propose that thermodynamics should be built into the underpinnings of a theory via an action principle; that Entropic Gravity compels a fully formulated theory of Relativistic Thermodynamics; that the perspective Entropic Gravity suggests possible methods of extracting Quantum Gravity predictions via an effective field theory, and that the emergence of Quantum Mechanics can be shown through the Statistical Mechanics of structures that resemble Markov-Chains.
	More work needs to be done in formalizing the details and construction of our above claims (such as unambiguously showing the complete emergence of Quantum Mechanics, extracting an actual Quantum Gravity  prediction from effective field theory, and a full exploration of the implications of our suggested formulation of Relativistic Thermodynamics), but we must remember that there is yet no definitive evidence suggesting that gravity is truly an entropic force. Though Entropic Gravity gives rise to some alluring calculations, a possible calculation of the cosmic dark energy density, this may be simply a coincidence. On what fronts progress should occur is not clear, as Verlinde's proposal doesn't immediately resolve issues physicists have been previously having, but it does strongly suggest certain areas for exploration. Specifically, the relationship between the emergence of gravity in a String theoretic context hints at an emergent Quantum Mechanics under String Theory's U, S, or T-Dualities. As Verlinde suggests that closed Strings be integrated out, it is clear that he does not discard String Theory entirely, so then the use of the U,S, and T-Dualities would still be entirely appropriate as a tool to analyze and check the self-consistency of Verlinde's proposal. If analysis of Verlinde's proposal under String Theory's dualities leads to a contradiction, then it may very well be all the evidence that is needed to show that gravity is not an entropic force. Conversely if it instead demonstrates how Quantum Mechanics can emerge from Statistical Mechanics and Thermodynamics, entirely untrodden avenues of exploration in physics may open up.
	 
		

\setcounter{secnumdepth}{0}

\section{Acknowledgements}

I would especially like to thank John Schwarz for mentoring my research this summer and serving as an excellent physics advisor. I would like to thank Sean Carroll, Erik Verlinde, Mark Wise, and Alexander Rakitin for their helpful and inspiring discussions, and I would like to thank Caltech's SURF program and the Ernest R. Robert's Fellowship for making this research possible.

\end{document}